\documentclass[twocolumn,showpacs,prl,preprintnumbers,amsmath,amssymb]{revtex4}
\usepackage{bm}
\usepackage{graphicx}
\usepackage{color}

\begin{document}

\title{Resonant light delay in GaN with ballistic and diffusive propagation}

\author{T.~V.~Shubina$^1$} 
\author{M.~M.~ Glazov$^1$}
\author{A.~A.~Toropov$^1$}
\author{N.~A.~Gippius$^2$}
\author{ A.~Vasson$^2$}
\author{ J.~Leymarie$^2$}
\author{A.~Kavokin$^3$}
\author{A.~Usui$^4$}
\author{J.~P.~Bergman$^5$}
\author{G.~Pozina$^5$}
\author{B.~Monemar$^5$}
\affiliation{$^1$Ioffe Physico-Technical Institute, Russian Academy of
Sciences, 194021 St. Petersburg, Russia}
\affiliation{$^2$LASMEA-UMR 6602 CNRS-UBP, 63177  Cedex, France }
\affiliation{$^3$University of Southampton, Highfield, Southampton, SO17 1BJ , United Kingdom}
\affiliation{$^4$R$\&$D Division, Furukawa Co., Ltd. Tsukuba, Ibaraki
305-0856, Japan}
\affiliation{$^5$Department of Physics, Chemistry and Biology,
Link\"oping University, S-581 83 Link\"oping, Sweden}

\begin{abstract}
We report on a strong delay in light propagation through bulk GaN, detected by time-of-flight spectroscopy. The delay increases resonantly as the photon energy approaches the energy of a neutral-donor bound exciton (BX), resulting in a
velocity of light as low as 2100 km/s. In the close vicinity of the BX resonance, the transmitted light contains both ballistic and diffusive components. This phenomenon is quantitatively explained in terms of optical dispersion in a medium where resonant light  scattering by the BX resonance takes place in addition to the polariton propagation.

\end{abstract}


\maketitle

Production of `slow light' is a subject of extensive studies nowadays \cite{slow}.
Propagation of an electromagnetic wave through a dielectric medium can be
considerably slowed down via ballistic or diffusive propagation,
due to either peculiarities of dispersion or multiple scattering acts, respectively. The ballistic light propagation
through the dispersive medium with an absorption resonance exhibits strong
retardation of the energy transport in the vicinity of the resonance \cite{Loudon},
so that the group velocity $v_g$ can be formally zero or even negative \cite{Garrett}. In semiconductors with exciton resonances, this effect is closely related to the exciton-polariton formation \cite{Pekar}.
The pulse propagation time agrees with that  calculated
from the polariton group velocity
in different semiconductor materials, namely, in
GaP:N around an isolated bound exciton (BX) line \cite{Chu}, CuCl near a free
exciton resonance \cite{Kuwata},  and ZnO in the BX vicinity \cite{Xiong}.

Light diffusion involving multiple photon scattering has been observed in
various disordered media, including dielectric micro-spheres, polymers, and ZnS
nanocrystals \cite{Kurita}. It is accompanied by intriguing phenomena, such as weak
localization and coherent backscattering of light \cite{Albada}. A strong slow-down
of light transfer due to diffusive motion of photons was discussed in,
e.g., ~\cite{Barabanenkov}. Resonant back-scattering of light has been observed in
quantum well structures \cite{Langbein}. On the other hand, in bulk semiconductors,
no experimental evidence for slow-light formation due to multiple scattering has
been reported so far, to our knowledge.

In this Letter we report the first observation of a strong delay of light
propagation in a high quality bulk GaN crystal, which amounts to 470 ps for a 1 mm
long way, corresponding to a light velocity of about 2100 km/s. We demonstrate a
resonant nature of this phenomenon and show that both light diffusion and retarded ballistic propagation of exciton-polaritons can contribute to the delay. Due to the wide use of GaN in modern optoelectronics, this effect is promising for practical applications.

\begin{figure} [t]
\includegraphics{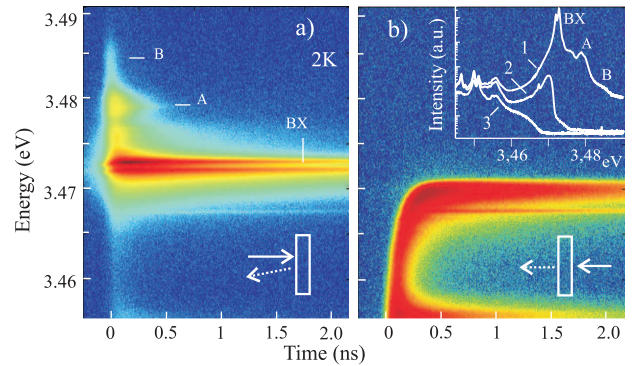}
  \caption{ \label{f1} (Color online) (a) Backward and (b) forward TR PL images recorded in the 1-mm sample with excitation by a 4.66 eV laser line. In the colored images, the PL intensity (logarithm scale) increases with the color variation from dark-blue to red. The inset presents time-integrated spectra of  backward PL in the 1-mm sample (1) and forward PL in the  1-mm (2) and  2-mm (3) samples.
   }
\end{figure}
The studies have been done by time-of-flight spectroscopy using free-standing GaN epilayers grown by HVPE, which are among the best available. They exhibit photoluminescence (PL) spectra with the width of neutral donor BX lines about 0.5 meV. We focused on three samples of different thicknesses, $L$: 1, 2 and 0.3 mm. The first sample has the carrier density of $8\times$10$^{15}$ cm$^{-3}$, two others around 5$\times$10$^{16}$ cm$^{-3}$. In these experiments, a Hamamatsu streak camera with a $\sim$20 ps temporal resolution was exploited to record the time-resolved (TR) images. The third and second harmonics from a tuned Ti:sapphire femtosecond pulsed laser were used as a light source.  The temporal width of the detected laser pulse (20-50 ps) was determined by the instrumental accuracy.  The transmitted light was registered in the forward direction along the  normal to the surface. Conventional backward PL was measured  for reference.

With excitation in a highly absorbing region by a 4.66 eV laser line, the backward
TR PL images (Fig. 1 a) show a fine excitonic structure with the pronounced A and B free exciton lines and the BX doublet at 3.472 eV, which is due to the exciton localization at Si and O donors. The transmitted PL images turn to be absolutely
different.  They exhibit a sharp quenching of the signal at a
boundary located just below this doublet  (Fig. 1 b).
This boundary at $\sim$3.47 eV is rather sharp in the best 1-mm sample;
in other samples it is more diffuse and red-shifted.
The other remarkable feature is the bending of the PL streak towards this boundary.
It means that the emission leaving the sample appears with a delay in time,
which increases with approaching the BX resonances.

\begin{figure} [t]
\includegraphics{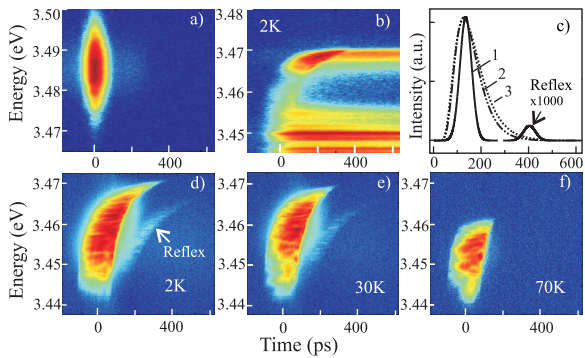}
  \caption{ \label{f2} (Color online) TR images (logarithm scale) of  3.486 eV pulses recorded (a) without a sample, (b) through the 1-mm sample. (c) The pulse shape at 3.465 eV calculated  for  ballistic (1) and diffusive  propagations in the sample  neglecting (2) and taking into account (3) the 97\% light reflection at boundaries. (d)-(f) TR images of 3.464 eV pulses passing through the sample at different temperatures.
   }
\end{figure}
When the laser pulse was tuned to 3.486 eV to be resonant with the free excitons, the transmitted signal includes the excited PL along with the non-absorbed part of the pulse.  Both components are cut and bent in a similar way (Fig. 2, b). The excited PL was negligible, when the excitation energy was tuned to 3.464 eV, i. e. below the exciton resonances. This permits us to study the propagation of the laser pulse solely.  Again, we observe the similar bending towards BX.  The images recorded at different temperatures show a shift of the cut-off with the increase of temperature.  This shift is $\sim$10 meV in the 2---70 K range, which is consistent with the variation of the band gap in GaN and, hence, of the BX energy.

There is one more important feature - an additional curved streak (``reflex''),
which is clearly seen in the images without PL taken with high resolution (Fig. 2 d-f). Its bending corresponds to the threefold delay time of the first transmitted pulse.
We assume that this
feature appears due to the triple coherent passage of a wave packet
(forward-back-forward) provided by the light reflection at the sample boundaries.
This seems to indicate that the pulse propagation in our samples is predominantly
ballistic, otherwise a photon should lose the direction memory after a few acts of scattering. The diffusion model predicts an extension of the pulse in the time domain but not an extra peak. This is illustrated by Fig. 2 (c), which presents the calculated pulse shapes, based on the models described below.

On the other hand, the reflex is clearly seen only in the best 1-mm
sample. In other samples with higher impurity concentration, it is less pronounced;
usually only a smooth cloud is recorded. Besides, the reflex is not visible close to
BX and disappears with the temperature rise, while the delay still exists. All this
shows that a non-ballistic contribution to the delay of light in the BX vicinity
cannot be excluded.

\begin{figure} [t]
\includegraphics{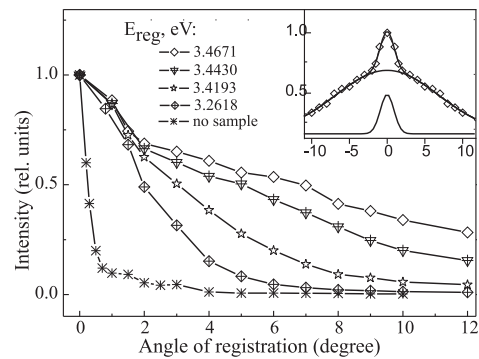}
  \caption{ \label{f3} Normalized angular diagrams of the light intensity measured at different photon energies in the 1-mm sample (and without that). The inset presents deconvolution of a diagram into narrow and broad components. The width of the ballistic part is determined by the light scattering at the sample surface. Some distortion of the broad component at large angles is due to the shading by cryostat windows.
   }
\end{figure}
In an attempt to gain further information, we have measured the far-field angular
dependence of the transmitted light. We expect the ballistic propagation to yield a
narrow angular diagram of the output intensity, while the diffusion is likely to
provide a much broader distribution. For this experiment, we used a Xe lamp, whose
light passed through a monochromator impinging normally to the sample surface. The
output intensity was registered at different angles by means of rotation of a stage,
holding the lamp and a cryostat, with respect to a detecting system.
As one can see from Fig. 3, the angular distribution of the light intensity
clearly indicates a superposition of two components - relatively narrow and broad
ones. We ascribe them to the ballistic and diffusive parts of the
transmitted light, respectively. We found that the intensity ratio between two components depends on the energy. The ballistic part dominates in the region of relative transparency, while the diffusive mechanism plays the major role in the vicinity of the BX resonances.

The coexistence of two regimes of light propagation is a surprising observation. It
is well-known that the ballistic propagation time $T_B(\omega)$ depends linearly on the sample
thickness $T_B(\omega)=L/v_g(\omega)$, while the delay time caused by a conventional diffusion
process would increase quadratically with the sample thickness. Thus the presence of the diffusive component could result in a marked extension of the streaks in the
time-resolved images, which is not observed. The question arises why so different
processes yield similar propagation times. To clarify that we model both processes separately.

\begin{figure} [t]
\includegraphics{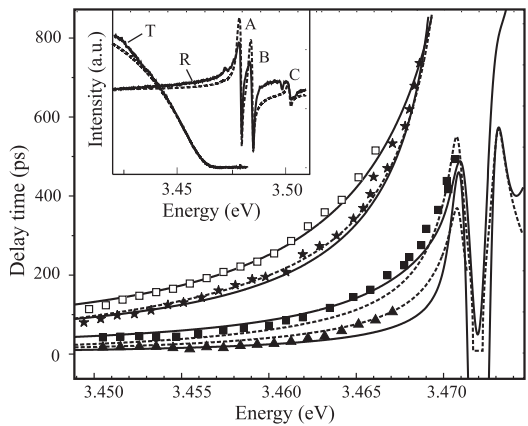}
  \caption{ \label{f4} Delay times measured in the 1-mm (squares), 2-mm (stars), and 0.3-mm (triangles) samples. Open squares are the data on the reflex in the 1-mm sample.  The fitting curves are presented for ballistic  (solid lines) and diffusive (dotted lines) propagation. The inset presents reflection (R) and transmission (T) spectra, measured using a lamp in the 1-mm sample, with their simulations (dotted lines).
   }
\end{figure}
For the ballistic propagation, the group velocity $v_g(\omega)=d\omega/dk$, where the wave vector $k(\omega)=(\omega/c)\sqrt{\varepsilon(\omega)}$. The frequency
dependent dielectric constant in the vicinity of exciton resonances can be written
neglecting spatial dispersion as
\begin{equation} \label{eq1}
\varepsilon(\omega)=\varepsilon_b+\sum_j\int\frac{f_j
\omega_{0,j}}{\omega_{0,j}+\xi-i\Gamma_j-\omega}
\frac{1}{\sqrt{\pi}\Delta_j}\exp{\left(\frac{-\xi^2}{{\Delta_j}^2}\right)}d\xi
\end{equation}
Here, $\varepsilon_b$ is the background dielectric constant, $j=A,B,C,BX$ denotes an
exciton resonance. For simplicity we consider only one bound state, making no
difference between an exciton localized at O or Si donors. Each resonance is
described by a frequency $\omega_{0,j}$, an oscillator strength $f_j$, and a damping
term $\Gamma_j$. The inhomogeneous broadening of the exciton line is taken into
account by means of convolution with the Gaussian with the
width $\Delta_{j}$. We underline that only
this formalism has allowed us to fit the reflection and transmission spectra
simultaneously (see inset in Fig. 4). Note that the conventional approach
\cite{ToriiK} yields a reasonable fitting of reflection spectra but strongly
overestimates the absorption of light near the exciton resonances.

We have performed a modeling assuming $\varepsilon_b=9.5$  and $\omega_{0,j}$ equal to 3.4720, 3.4785, 3.4837, and 3.5016 eV for BX, A, B, and C  excitons, respectively. For either of the resonance, $\Gamma_j$  is taken as 0.012 meV, which is of the order of the homogeneous line width of an exciton-polariton in other semiconductors \cite{Masumoto}. The inhomogeneous width $\Delta_j$ is taken as 0.75 meV, close to the average PL linewidth in our spectra. For the A, B and C excitons, $f$ equals to 0.0026, 0.0017, and 0.00037, respectively, which is comparable with the previously reported values for GaN \cite{ToriiK}.

The oscillator strength for the bound exciton with the donor concentration $N$ can be estimated as~\cite{Ivchenko}
\begin{equation}\label{eq2}
f= \frac{4\pi N}{\hbar \omega_{0,BX}} \left(\frac{e|p_{cv}|}{m_0\omega_{0,BX}}\right)^2  \left(\int \Psi(\bm r, 0) d\bm r\right)^2,
\end{equation}
where $m_0$ is the free electron mass, $e$ is the elementary charge, $p_{cv}$ is the interband matrix element, and $\Psi(\bm r, \bm \rho)$ is the envelope of the electron-hole pair function depending on the center of mass position $\bm r$ and the relative electron-hole vector $\bm \rho$. Under the assumption that the exciton is localized as a whole it has the form
\begin{equation}\label{eq:psi}
\Psi(\bm r,\bm \rho) = \frac{1}{\pi R^{3/2}a_B^{3/2}} \exp{(-\bm{r}/R)} \exp{(-\bm{\rho}/a_B)}.
\end{equation}
Here we took the exciton localization radius $R={a_B}$, where $a_B=2.8$ nm is the Bohr radius, which allows us to reproduce the observed value of the BX binding energy ~\cite{Suffcszynski}.  $N$ is taken equal to the carrier density; $p_{cv}=10^{-19}$ g cm$/$s. The best fit is obtained with $f\approx2\times 10^{-5}$ for the 1-mm sample, what is somewhat smaller than an estimation according to Eq.  \eqref{eq2}. For other samples f is taken an order of magnitude higher following the
difference in the carrier densities. Figure 4 demonstrates a reasonable fit with these parameters and clarifies the origin of the sharp cut-off. The calculated dependencies have a strong dip near the BX resonance.  It appears because the phase and group velocity of light strongly decrease here, while the absorption is enhanced~\cite{Garrett}.

To consider the diffusive propagation we assume that elastic neutral donor scattering dominates the delay, neglecting any inelastic scattering by impurities or acoustic phonons \cite{Steiner}.  We write the diffusion equation for the concentration of photons $n(\omega,x,t)$ at frequency  $\omega$, coordinate $x$, and time $t$
\begin{equation}\label{eq3}
\frac{\partial n(\omega,x,t)}{\partial t} = D(\omega)\frac{\partial^2
n(\omega,x,t)}{\partial x^2} - \frac{n(\omega,x,t)}{\tau(\omega)}.
\end{equation}
Here frequency dependent $D(\omega)$ and $\tau(\omega)$ are a  diffusion coefficient and
photon life-time, respectively.  The life-time is defined as $\tau(\omega)=\tau_0\cdot(l_A(\omega)/l_0)$, where  $l_0$ and $\tau_0$ are a photon mean free path and time between two acts of scattering in the region of relative transparency; $l_A(\omega)$ is the absorption length.  All the photons are assumed to be initially created in a thin layer near the sample surface: $n(\omega,x,0=n(\omega,0)\delta (x).$
We have performed calculations of the pulse propagation in the diffusive model neglecting and taking into account the light reflection at boundaries  and found no qualitative difference (Fig. 2, c). Below the light reflection is neglected.
The delay time due to diffusive photon motion, $T_D(\omega)$, can be found from the condition ${\partial n(\omega,L,t)}/{\partial t} = 0$ which corresponds to the intensity maximum. In the general case it is written as
\begin{equation}\label{eq4}
T_D(\omega) = \sqrt{\frac{\tau(\omega)^2}{16} + \frac{L^2\tau(\omega)}{4D(\omega)}}
- \frac{\tau(\omega)}{4}.
\end{equation}
We have the two following limiting cases: i) small absorption, $\tau(\omega) \gg L^2/D(\omega)$ where the delay scales as a second power of the sample thickness, $T_D(\omega) = {L^2}/{[2D(\omega)]}$; ii) strong absorption, $\tau(\omega) \ll  L^2/D(\omega)$, where the delay is proportional to the sample thickness as with the ballistic propagation, $T_D(\omega) = ({L}/{2})\sqrt{{\tau(\omega)}/{D(\omega)}}$. The latter result is a consequence of the competition between the `fast' absorption and `slow' diffusion processes.

The diffusion constant $D(\omega)=v_g(\omega) l_{tr}(\omega)/3$, where $l_{tr}(\omega)=(N\sigma(\omega))^{-1}$ is the photon mean free path determined by the light scattering cross-section $\sigma(\omega)$ and the concentration of scattering centers $N$. Assuming that the light scattering is caused by the bound exciton resonances one may write $\sigma(\omega)$ in a general form~\cite{landauIV}~\footnote{Our estimations show that the contribution to Eq. \eqref{eq5} from the light scattering by free excitons does not exceed $20$\%. For the purposes of the paper, we neglect that as well as the BX resonance inhomogeneous broadening.}
\begin{equation}\label{eq5}
\sigma(\omega) =  {\frac{f^2}{4\pi N^2 c^{4}}}   {\frac{\omega_{0,BX}^{6}}{(\omega_{0,BX}-\omega)^{2}+\gamma^2}}.
\end{equation}
Here, $\gamma$ is an effective broadening parameter. The main contribution to the delay time comes from the strong frequency dependence of the cross-section. Strictly speaking, Eq.  \eqref{eq5} describes scattering caused by individual BX. This is valid only for the diluted donor concentration satisfying the condition $N^{1/3}\lambda \ll 1$, where $\lambda$ is a wavelength  inside the crystal. Near BX, this is fulfilled for $N\sim$10$^{15}$ cm$^{-3}$, while the donor density is higher in our samples.  On the other hand, the spatial distribution of donors is inevitably inhomogeneous and the clusters of closely lying BX act as virtual scattering centers with effective concentration and localization radius. For the sake of demonstration, we present here $T_D(\omega)$  calculated as for the individual scatterers. We use $\gamma=0.75$~meV, $\tau_0/l_0=2\times$10$^{-9}$ s/cm, and $f$ calculated according to Eqs. \eqref{eq2}, \eqref{eq:psi}. While for the 1-mm sample some agreement is achievable with $N=8\times10^{15}$ cm$^{-3}$, for other samples we are impelled to take the effective $N$=2$\times$10$^{16}$ cm$^{-3}$ to obtain a reasonable fitting (Fig. 4).

Our simulations of the experimental data are done in the limiting cases of
either ballistic or diffusive light propagation. We abstain from the fitting in a
combined model; it hardly can be representative, because the ratio between these
components is too dependent on the sample quality. Our main goal is to demonstrate
that the diffusive mechanism can give the delay of the same order of magnitude as
the ballistic one, and that its dependence of the delay on the thickness can be
linear as well.

In conclusion, we reported here a strong delay in light transfer through high
quality bulk GaN samples and present a quantitative description of this phenomenon
in the frameworks of optical dispersion and resonant elastic scattering models. We
believe that our findings have a particular importance for the development of a new
generation of smart optoelectronic devices, which rely on a controlled delay of
light propagation.

\section*{Acknowledgments}
We thank Profs. E. L. Ivchenko and M. I. Dyakonov for fruitful discussions. This work is supported in part by the RFBR and ANR grants. A.A.T. acknowledges Wenner-Gren and Russian Science Support Foundations, M.M.G. is grateful to Dynasty Foundation - ICFPM. We appreciate the supply of samples by Prof. B. Beaumont.

\end{document}